%% file: kacprzak.tex
\documentclass[graybox, envcountchap, authoryear, natbib]{svmult}

\usepackage{mathptmx}       
\usepackage{helvet}         
\usepackage{courier}        
\usepackage{type1cm}        
 \usepackage{natbib}

\usepackage{makeidx}         
\usepackage{graphicx}        
\usepackage{multicol}        
\usepackage[bottom]{footmisc}

\newcommand{\CIV}{\hbox{{\rm C}\kern 0.1em{\sc iv}}}
\newcommand{\CV}{\hbox{{\rm C}\kern 0.1em{\sc v}}}
\newcommand{\HI}{\hbox{{\rm H}\kern 0.1em{\sc i}}}
\newcommand{\HII}{\hbox{{\rm H}\kern 0.1em{\sc ii}}}
\newcommand{\Lya}{\hbox{{\rm Ly}\kern 0.1em$\alpha$}}
\newcommand{\Lyb}{\hbox{{\rm Ly}\kern 0.1em$\beta$}}
\newcommand{\Lyg}{\hbox{{\rm Ly}\kern 0.1em$\gamma$}}
\newcommand{\Lyd}{\hbox{{\rm Ly}\kern 0.1em$\delta$}}
\newcommand{\Lye}{\hbox{{\rm Ly}\kern 0.1em$\epsilon$}}
\newcommand{\Lyphi}{\hbox{{\rm Ly}\kern 0.1em$\phi$}}
\newcommand{\Lyfive}{\hbox{{\rm Ly}\kern 0.1em$5$}}
\newcommand{\Lysix}{\hbox{{\rm Ly}\kern 0.1em$6$}}
\newcommand{\Lyseven}{\hbox{{\rm Ly}\kern 0.1em$7$}}
\newcommand{\Lyeight}{\hbox{{\rm Ly}\kern 0.1em$8$}}
\newcommand{\Lynine}{\hbox{{\rm Ly}\kern 0.1em$9$}}
\newcommand{\Lyten}{\hbox{{\rm Ly}\kern 0.1em$10$}}
\newcommand{\Lyeleven}{\hbox{{\rm Ly}\kern 0.1em$11$}}
\newcommand{\HeI}{\hbox{{\rm He}\kern 0.1em{\sc i}}}
\newcommand{\HeII}{\hbox{{\rm He}\kern 0.1em{\sc ii}}}
\newcommand{\FeI}{\hbox{{\rm Fe}\kern 0.1em{\sc i}}}
\newcommand{\FeII}{\hbox{{\rm Fe}\kern 0.1em{\sc ii}}}
\newcommand{\FeIII}{\hbox{{\rm Fe}\kern 0.1em{\sc iii}}}
\newcommand{\MnII}{\hbox{{\rm Mn}\kern 0.1em{\sc ii}}}
\newcommand{\MgI}{\hbox{{\rm Mg}\kern 0.1em{\sc i}}}
\newcommand{\MgIb}{\hbox{{\rm Mg}\kern 0.1em{\sc i}}\kern 0.05em{\rm b}}
\newcommand{\MgII}{\hbox{{\rm Mg}\kern 0.1em{\sc ii}}}
\newcommand{\MgIII}{\hbox{{\rm Mg}\kern 0.1em{\sc iii}}}
\newcommand{\NI}{\hbox{{\rm N}\kern 0.1em{\sc i}}}
\newcommand{\NII}{\hbox{{\rm N}\kern 0.1em{\sc ii}}}
\newcommand{\NIII}{\hbox{{\rm N}\kern 0.1em{\sc iii}}}
\newcommand{\NV}{\hbox{{\rm N}\kern 0.1em{\sc v}}}
\newcommand{\OVI}{\hbox{{\rm O}\kern 0.1em{\sc vi}}}
\newcommand{\OI}{\hbox{{\rm O}\kern 0.1em{\sc i}}}
\newcommand{\OII}{\hbox{[{\rm O}\kern 0.1em{\sc ii}]}}
\newcommand{\OIII}{\hbox{[{\rm O}\kern 0.1em{\sc iii}]}}
\newcommand{\OIV}{\hbox{{\rm O}\kern 0.1em{\sc iv}]}}
\newcommand{\SI}{{\rm S}\kern 0.1em{\sc i}}
\newcommand{\SIV}{{\rm S}\kern 0.1em{\sc iv}}
\newcommand{\SVI}{{\rm S}\kern 0.1em{\sc vi}}
\newcommand{\SiI}{\hbox{{\rm Si}\kern 0.1em{\sc i}}}
\newcommand{\SiII}{\hbox{{\rm Si}\kern 0.1em{\sc ii}}}
\newcommand{\SiIII}{\hbox{{\rm Si}\kern 0.1em{\sc iii}}}
\newcommand{\SiIV}{\hbox{{\rm Si}\kern 0.1em{\sc iv}}}
\newcommand{\SII}{\hbox{{\rm S}\kern 0.1em{\sc ii}}}
\newcommand{\SIII}{\hbox{{\rm S}\kern 0.1em{\sc iii}}}
\newcommand{\NaI}{\hbox{{\rm Na}\kern 0.1em{\sc i}}}
\newcommand{\NaID}{\hbox{{\rm Na}\kern 0.1em{\sc i}}\kern 0.05em{\rm D}}
\newcommand{\TiII}{\hbox{{\rm Ti}\kern 0.1em{\sc ii}}}
\newcommand{\kms}{\hbox{~km~s$^{-1}$}}

\begin{document}

\title*{Gas Accretion in Star-Forming Galaxies}
\author{Glenn G. Kacprzak}
\institute{ Glenn G. Kacprzak \at Swinburne University of Technology
Center for Astrophysics \& Supercomputing, Victoria 3122, Australia,
  \email{gkacprzak@swin.edu.au}}
%
%
\maketitle

\abstract*{Cold-mode gas accretion onto galaxies is a direct
  prediction of $\Lambda$CDM simulations and provides galaxies
  with fuel that allows them to continue to form stars over the
  lifetime of the Universe.  Given its dramatic influence on a
  galaxy's gas reservoir, gas accretion has to be largely responsible
  for how galaxies form and evolve.  Therefore, given the importance
  of gas accretion, it is necessary to observe and quantify how these
  gas flows affect galaxy evolution. However, observational data have
  yet to conclusively show that gas accretion ubiquitously occurs at
  any epoch. Directly detecting gas accretion is a challenging
  endeavor and we now have obtained a significant amount of
  observational evidence to support it. This chapter reviews the
  current observational evidence of gas accretion onto star-forming
  galaxies.}

\abstract{Cold-mode gas accretion onto galaxies is a direct prediction
  of $\Lambda$CDM simulations and provides galaxies with fuel
  that allows them to continue to form stars over the lifetime of the
  Universe.  Given its dramatic influence on a galaxy's gas reservoir,
  gas accretion has to be largely responsible for how galaxies form
  and evolve.  Therefore, given the importance of gas accretion, it is
  necessary to observe and quantify how these gas flows affect galaxy
  evolution. However, observational data have yet to conclusively show
  that gas accretion ubiquitously occurs at any epoch. Directly
  detecting gas accretion is a challenging endeavor and we now have
  obtained a significant amount of observational evidence to support
  it. This chapter reviews the current observational evidence of gas
  accretion onto star-forming galaxies.}



\section{Introduction}
\label{sec:0}

Cosmological simulations unequivocally predict that cold accretion is
the primary growth mechanism of galaxies. The accretion of cold gas
occurs via cosmic filaments that transports metal-poor gas onto
galaxies, providing their fuel to form stars. Observationally, it is
quite clear that galaxy gas-consumption timescales are short compared
to the age of the Universe, therefore galaxies must acquire gas from
the surroundings to continue to form stars. However, observational
data have yet to conclusively show that gas accretion flows are
ubiquitously occurring in-and-around galaxies at any epoch.  Since the
first discoveries of circumgalactic gas around star-forming galaxies
\citep{boksenberg78,kunth84,bergeron86}, we have wondered where does this gas
come from and if/how it drives galaxy growth and evolution.

This Chapter reviews the current observational evidence and signatures
of cold accretion onto star-forming galaxies. In the following
sections, we review the data that suggest accretion is occurring and
cover the main topics of circumgalactic gas spatial distribution,
kinematics and metallicity. In a few cases, there exists the
combination of all of the above which provides the most tantalizing
evidence of cold accretion to date. This Chapter primarily focuses on
observations using background quasars or galaxies as probes of the
circumgalactic medium around intervening foreground galaxies.

\section{The Spatial Distribution of the Circumgalactic Medium}
\label{sec:1}

In order for galaxies to continuously form stars throughout the age of
the Universe, they must acquire a sufficient amount of gas from their
surroundings. In fact, roughly 50\% of a galaxy's gas mass is in the
circumgalactic medium \citep{zheng15,wolfe05}. Using background
quasars as probes of gas surrounding foreground galaxies, we have
discovered that galaxies have an abundance of multi-phased
circumgalactic gas.

\subsection{Circumgalactic Gas Radial Distribution}

\begin{figure}[h]
\includegraphics[scale=0.89]{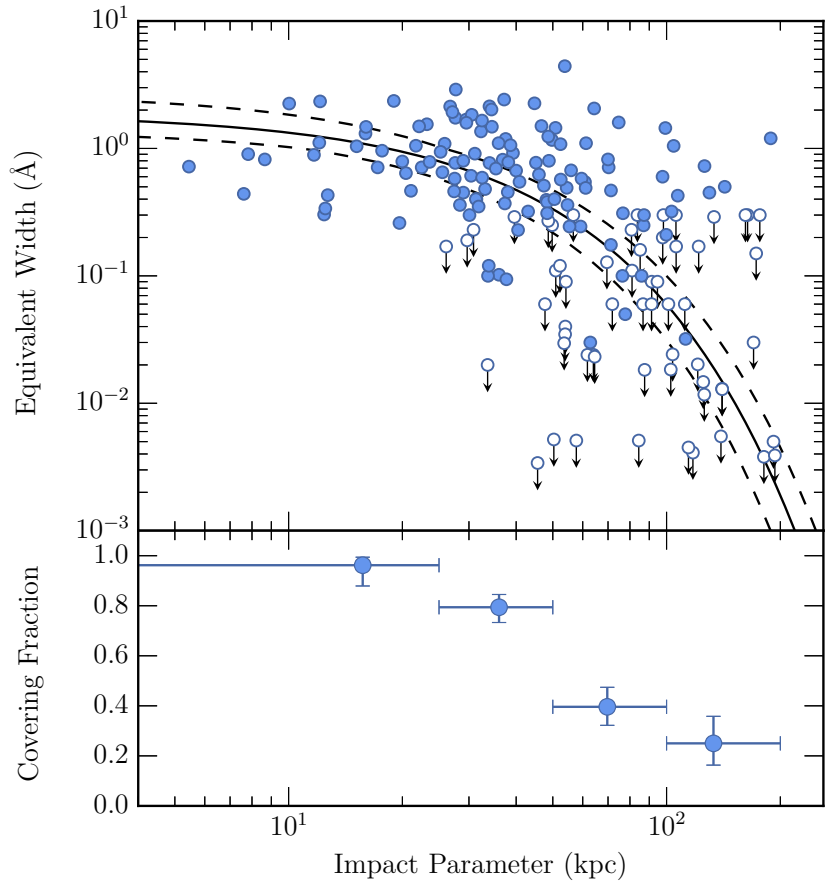}
\caption{{\it Top}: the rest-frame {\MgII} equivalent width as a function
  of impact parameter for the MAGIICAT catalog for ``isolated''
  galaxies \citep{nielsen13a,nielsen13b}. The closed symbols are
  detections while open symbols are 3$\sigma$ are upper limits. The
  fit, and 1$\sigma$ confidence levels are shown (log$W_r(2796) =
  [−0.015\pm0.002]\times D+ [0.27\pm0.11$]).  Note the large extent of
  gas surrounding galaxies, which begs the question of what is the
  origin of this gas and is it some combination of gas outflows and
  accretion. {\it Bottom}: the radial decline of the gas covering
  fraction as a function of impact parameter for an equivalent
  detection limit of 0.3~{\AA}. Image courtesy of Nikole Nielsen.}
\label{fig:1ewd}       
\end{figure}

The quantity and extend of the circumgalactic medium has been traced
using a range of absorption features such as {\Lya}
\citep{tripp98,chen01b,wakker09,chen09,steidel10,stocke13,richter16},
{\MgII} \citep{sdp94,s95,gb97,zibetti07,kacprzak08,chen10,nielsen13a},
{\CIV}
\citep{chen01a,adelberger05,steidel10,bordoloi14a,liang14,burchett15,richter16},
and {\OVI} \citep{savage03,sembach04,stocke06, danforth08,
  tripp08,wakker09,prochaska11,tumlinson11,johnson13,stocke13,johnson15,kacprzak15}.
These studies all have shown that regardless of redshift (at least
between $z=0-3$), galaxies typically have hydrogen gas detected out to
$\sim$500 kpc with ``metal-enriched'' gas within 100--200 kpc.
Furthermore, the data show an anti-correlation with equivalent width
and impact parameter, with the covering fraction being unity close to
the galaxy and declining with increasing distance.  This is
demonstrated in Figure~\ref{fig:1ewd}, which shows the $\sim$200~kpc
extent of {\MgII} absorbing gas around typical galaxies from the
MAGIICAT catalog \citep{nielsen13b} along with well fit
anti-correlation between equivalent width and impact parameter
(log$W_r(2796) = −0.015\times D+ 0.27$). Note also that the gas
covering fraction, for an equivalent width limit of 0.3~{\AA}, is
roughly unity near the galaxy and decreases to about 20\% beyond
100~kpc.

Interestingly, \citet{richter12} demonstrated, using the distribution
of high-velocity clouds around the Milky Way and M31, that
high-velocity clouds could give rise to the majority of the absorption
systems seen around other galaxies.  The accretion rate of
high-velocity gas at $z=0$ is almost equivalent to the star formation
density of the local Universe and thus, at least at low redshifts,
high-velocity clouds could provide a significant fraction of the gas
mass accreted onto galaxies (see chapter by Philipp Richter for further discussion of
gas accretion onto the Milky Way).

Simulations predict that gas accretion should occur via a ``hot" or
``cold" mode, which is dependent on a galaxy being above or below a
critical halo mass ranging between log(M$_h$)=11--12
\citep{birnboim03,keres05,dekel06,ocvirk08,brooks09,dekel09,keres09,stewart11,vandevoort11}. A
repercussion of these models is that the covering fraction of cool
accreting gas should drop significantly to almost zero for massive
galaxies \citep{stewart11}. However, observational evidence shows that
the covering fraction of cold gas is constant over a larger range of
halo masses $10.7 \leq$~log(M$_h$)~$\leq 13.9$ within a given impact
parameter (or impact parameter normalized by the virial radius) and
that gaseous galaxy halos are self-similar
\citep{churchill13a,churchill13b}. This is suggestive that either
outflows and/or other substructures contribute to absorption in
high-mass halos such that low- and high-mass gas halos are
observationally indistinguishable or the data indicate that
predictions of a mass dependent shutdown of cold-mode accretion may
require revision. This area needs to be examined further in order to
address the discrepancy between observations and simulations.

Although the average radial gas profiles around star-forming galaxies
are well quantified, it does not provide much insight into the nature
of the circumgalactic gas. Thus, it is critical to determine the
geometric distribution of gas relative to its host galaxy to help
improved our understanding of its origins.

\subsection{Circumgalactic Gas Spatial Distribution}


In the mid-90s, the exploration of circumgalactic medium geometry
started with \citet{s95} who acquired a large sample of 51
galaxy-quasar pairs. The data were suggestive that, independent of
galaxy spectroscopic/morphological type, {\MgII} gas resided within a
spherical halo with unity gas covering fraction. The data fit well to
a Holmberg-like luminosity scaling between a characteristic halo
radius and galaxy K-band luminosity. However, even Steidel noted that
spherical halos were likely a tremendous over-simplification of the
true situation, however, the data did not disprove it. Using simple
geometric models, \citet{charlton96} determined that both spherical
halos and extended monolithic thick-disk models could be made
consistent with the current data. They suggested that the kinematic
structure of the absorption profiles could be used to further
constrain the gas geometry, which we further discuss in
Section~\ref{sec:2}.

Cosmological simulations commonly show that gas accretion should occur
along filaments that are co-planar to the galaxy disk, whereas gas
outflows are expected to be expelled along the galaxy projected minor
axis \citep[e.g.,][]{shen12,stewart16}.  Reminiscent of
\citet{charlton96}, \citet{kacprzak11b} reported that the {\MgII}
equivalent width measured from high resolution quasar spectra was
dependent on galaxy inclination, suggesting that the circumgalactic
medium has a co-planer geometry that is coupled to the galaxy
inclination. It was noted however that the absorbing gas could arise
from tidal streams, satellites, filaments, etc., which could also have
somewhat co-planer distributions. By stacking over 5000 background
galaxies to probe over 4000 foreground galaxies, \citet{bordoloi11}
found a strong azimuthal dependence of the {\MgII} absorption within
50 kpc of inclined disk-dominated galaxies \citep[also
see][]{lan14}. They found elevated equivalent width along the galaxy
minor axis and lower equivalent width along the major axis. Their data
are indicative of bipolar outflows with possible flows along the major
axis.  Later, \citet{bordoloi14b} presented models of the
circumgalactic medium azimuthal angle distribution by using joint
constraints from: the integrated {\MgII} absorption from stacked
background galaxy spectra \citep{bordoloi11} and {\MgII} absorption
from individual galaxies as seen from background quasar spectra
\citep{kacprzak11b}.  They determined that either composite models
consisting of a bipolar outflow component plus a spherical or disk
component, or a single highly softened bipolar distribution, could
well represent data within 40 kpc.

\begin{figure}[b]
\includegraphics[scale=0.85]{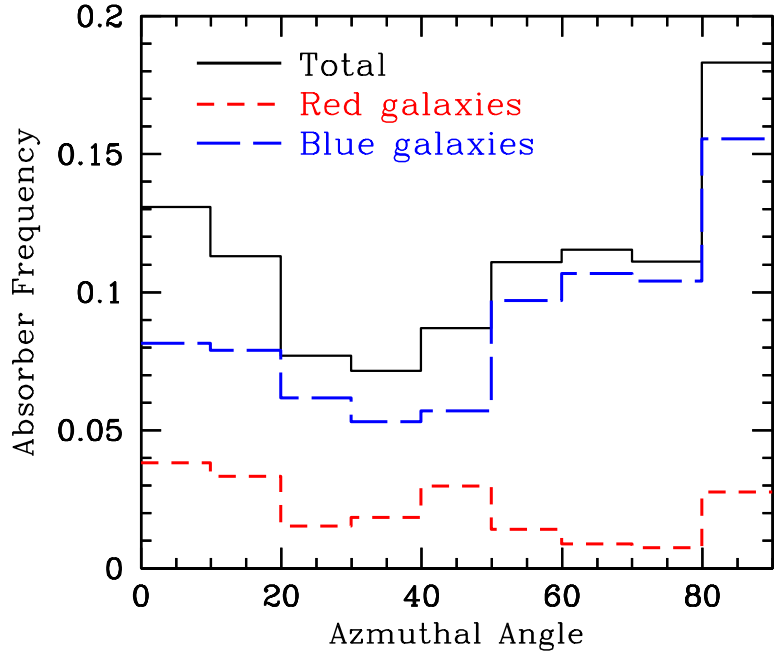}
\caption{Binned azimuthal angle mean probability distribution function
  (PDF) for {\MgII} absorbing galaxies (solid line). The binned PDFs are
  normalized such that the total area is equal to unity, yielding an
  observed frequency in each azimuthal bin. Absorption is detected
  with increased frequency toward the major ($\Phi=0$~degrees) and
  minor ($\Phi=90$~degrees) axes. Also shown is the galaxy color
  dependence of the distribution split by a $B-R\leq1.1$
  representing late-type galaxies (dashed blue line) and $B-R > 1.1$
  representing early-type galaxies (dotted red line).  The data are
  consistent with star-forming galaxies accreting gas and producing
  large-scale outflows while quiescent galaxies have much less gas
  activity. }
\label{fig:1}       
\end{figure}

\citet{bouche12}, using 10 galaxies, first showed that the azimuthal
angle distribution of absorbing gas traced by {\MgII} appeared to be
bimodal with half of the {\MgII} sight-lines showing a co-planar
geometry.  \citet{kacprzak12a} further confirmed the bimodality in the
azimuthal angle distribution of gas around galaxies, where cool dense
circumgalactic gas prefers to exist along the projected galaxy major
and minor axes where the gas covering fraction are enhanced by
20\%--30\% as shown in Figure~\ref{fig:1}.  Also shown in
Figure~\ref{fig:1} is that blue star-forming galaxies drive the
bimodality while red passive galaxies may contain gas along their
projected major axis. The lower equivalent width detected along the
projected major axis is suggestive that accretion would likely contain
metal poor gas with moderate velocity width profiles.

The aforementioned results provide a geometric picture that is
consistent with galaxy evolution scenarios where star-forming galaxies
accrete co-planer gas within a narrow streams with opening angles of
about 40 degrees, providing fuel for new stars that produce
metal-enriched galactic scale outflows with wide opening angles of 100
degrees, while red galaxies exist passively due to reduced gas
reservoirs. These conclusions are based on {\MgII} observations,
however both infalling gas and outflowing are expected to contain
multi-phased gas.

\citet{mathes14} first attempted to address the azimuthal angle
dependence for highly ionized gas traced by {\OVI} and found it to
have a spatially uniform distribution out to 300~kpc. Using a larger
sample of {\OVI} absorption selected galaxies, \citet{kacprzak15}
reported a bimodality in the azimuthal angle distribution of gas
around galaxies within 200~kpc.  Similar to {\MgII}, they found that
{\OVI} is commonly detected within opening angles of 20--40 degrees of
the galaxy projected major axis and within opening angles of at least
60 degrees along the projected minor axis. Again similar to {\MgII},
weaker equivalent width systems tend to reside on along the project
major axis. This would be expected for either lower column density,
lower kinematic dispersion or low metallicity (or a combination
thereof) gas accreting towards the galaxy major axis. Different from
the {\MgII} results, non-detections of {\OVI} exist almost exclusively
between 20--60 degrees, suggesting that {\OVI} is not mixed throughout
the circumgalactic medium and remains confined within the accretion
filaments and the gas outflows.

Further supporting this bimodality accretion/outflow picture is the
recent work using {\HI} 21-cm absorption to probe the circumgalactic
medium within impact parameters of $<35$~kpc around $z<0.4$ galaxies.
\citep{dutta16} found that the majority of their absorbers (nine)
exists along the projected major axis and a few (three) exists along
the projected minor axis. The data are supportive of high column
density co-planer {\HI} thick-disk around these galaxies. In addition,
the three minor axis absorption systems all reside within 15 kpc,
therefore they conclude that these low impact parameter minor axis
systems could originate from warps in these thick and extended {\HI}
disks.

Although gas geometry is highly suggestive of (and consistent with)
our exception of gas accretion onto star-forming galaxies, it alone is
not sufficient enough to determine if gas is actually fact accreting
onto galaxies. Gas and relative gas-galaxy kinematics can provide
additional data that can be used to address whether we are detecting
gas accretion or not.

\section{Circumgalactic Gas Kinematics}
\label{sec:2}

Absorption systems produced by the circumgalactic medium hold key
kinematic signatures into unlocking the the behavior of gas around
galaxies. High resolution spectroscopy of the background quasars is
critical to resolving the velocity substructures within these complex
absorption systems. These data can be used to differentiate between
scenarios of gas accretion, disk-rotation and outflows.

\subsection{Internal or Intrinsic Gas Kinematics}
\label{subsec:2.1}


For the most part, metal-line absorption systems are not composed of a
uniform velocity distribution of ``clouds'', but tend to exist in
groupings closer together in velocity with occasional higher velocity
clouds offset from the groupings.  For a handful high-resolution
{\MgII} absorption profiles, \citet{lanzetta92} inferred that their
velocity structure was dominated by coherent motions as oppose to
random. They further showed that the absorption profiles are
consistent with a rotating ensembles of clouds similar to a
co-rotating disk. With a larger sample of high-resolution absorption
profiles, \citet{charlton98} applied statistical tests for a variety
of kinematic models and concluded that pure disk rotation and pure
accretion models are likely ruled out. However, models with
contributions from both a rotating disk and infall/halo can reproduce
velocities that are nearly consistent with the observed kinematics.

Similar work focusing on the low-ion transitions (such as {\SiII})
associated with damped {\Lya} systems, \citet{prochaska97}
examined if a range of models such as rotating cold disks, slowly
rotating hot disks, massive isothermal halos, and a hydrodynamic
spherical accretion models could explain the observed absorption
kinematics.  They determined that thick rapidly rotating disks are the
only model consistent with the data at high confidence levels.  Their
tests suggest that disk rotation speeds of around 225~{\kms} are
preferred, which is typical for Milky Way-like galaxy rotation
speeds. Furthermore, the gas is likely to be cold since ratio of the
gas velocity dispersion over the disk rotation speed must be less than
0.1. These data are suggestive of thick, cold and possibly accreting
disks surrounding galaxies. 

All these studies are based on absorption-line data alone and
therefore it is important to quantify how the absorption kinematics
changes with galaxy properties.

Using primarily Lyman limits systems, \citet{borthakur15} found {\Lya}
absorption from the circumgalactic medium has similar velocity spreads
to that of their host galaxy's interstellar medium as observed via
{\HI} emission.  The combination of the correlation between the galaxy
gas fraction and the impact-parameter-corrected {\Lya} equivalent
width is consistent with idea that the {\HI} disk is fed by
circumgalactic gas accretion \citep{borthakur15}. Furthermore, they
find a correlation between impact-parameter-corrected {\Lya}
equivalent width and the galaxy specific star formation rate
suggesting a link between gas accretion driving star formation
\citep{borthakur16}.

\begin{figure}
\includegraphics[scale=0.59]{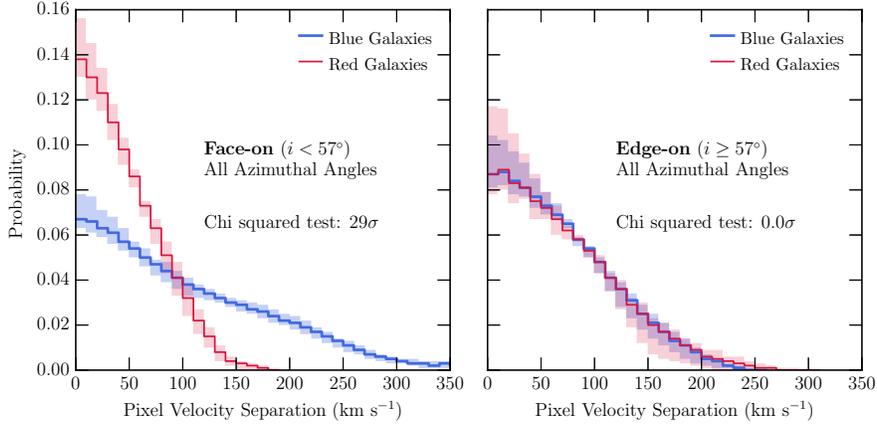}
%
%
\caption{Shown are pixel-velocity two-point correlation functions
  (TPCFs) examining how the spread in the pixel velocities of
  absorption profiles differ for different galaxy properties. The
  TPCFs presented give the probability of having pixel velocities at a
  given velocity offset from the absorption redshift defined by the
  optical depth weighed mean.  The shaded regions are the 1$\sigma$
  bootstrap uncertainties.  For all azimuthal angles, the TPCFs are
  shown for blue and red face-on galaxies (left) and edge-on galaxies
  (right).  Note the dramatic differences between the absorber
  velocity dispersions for blue and red galaxies for face-on
  orientations (left). The larger velocity spread is suggestive of of
  high velocity outflows being ejected from the galaxy, while red
  galaxies are less active.  On the other-hand, there is no difference
  between blue and red for edge-on orientations (right). The velocity
  spread is comparable to the rotation speed of galaxies, which is 
  consistent with gas accretion and/or gaseous disk co-rotation around
  the galaxies.}
\label{fig:tpcf}       
\end{figure}

\citet{nielsen15} has quantified the {\MgII} absorber velocity
profiles using pixel-velocity two-point correlation functions and
determined how absorption kinematics vary as a function of the galaxy
orientation and other physical properties as shown in
Figure~\ref{fig:tpcf}. While they find that absorption profiles with
the largest velocity dispersion are associated with blue, face-on
galaxies probed along the projected minor axis, which is suggestive
out outflows, they find something different for edge-on galaxies.  For
edge-on galaxies probed along the major axis, they find large {\MgII}
absorber velocity dispersions and large column density clouds at low
velocity regardless of galaxy color, which is used as a tracer of star
formation rate. It is suggested that the large absorber velocity
dispersions seen for edge-on galaxies (Figure~\ref{fig:tpcf} -- right)
may be caused by gas rotation/accretion where the line-of-sight
velocity is maximized for edge-on galaxy inclination. Furthermore, the
large cloud column densities may indicate that co-rotating or
accreting gas is fairly coherent along the line-of-sight. The only way
to test this scenario is to compare the absorption velocities relative
to the rotation velocity of the host galaxies.

In addition, some evolution in the circumgalactic medium has also been
observed. Blue galaxies do not show an evolution in the velocity
dispersions and cloud column densities with redshift (between $0.3\leq
z \leq 1$), while red galaxies have a circumgalactic medium that
becomes more kinematically quiescent with time \citep{nielsen16}. This
is suggestive that the gas cycle in blue star-forming galaxies is
active, be it via accretion or outflows, while red galaxies exhibit
little-to-no gas activity. This is consistent with the little-to-no
{\OVI} found around $z\sim 0.25$ quiescent galaxies from the
COS$-$Halos survey \citep{tumlinson11}.

\subsection{Relative Gas-Galaxy Kinematics}
\label{subsec:2.2}

It is interesting that the internal velocity structure of absorption
systems are reflective of their host galaxy type, orientation and
redshift, however the question then arises of how/if the
circumgalactic medium is kinematically connected to their host
galaxies.  The most direct measure of gas accretion is observing it
down-the-barrel by using the host galaxy as the background
source. This method is ideal since there are no degeneracies in the
line-of-sight direction/velocity. Although this method does have its
difficulties too since metal-enriched outflows and the interstellar
medium will tend to dominate the observed absorption over the
metal-poor accreting gas.  These down-the-barrel gas accretion events
have has been observed in a few cases with absorption velocity shifts
relative to the galaxy of $80-200$~{\kms}
\citep{martin12,rubin12}. These down-the-barrel gas accretion
observations are discussed in detail in the chapter by Kate Rubin. It is still unknown
if these observations are signatures of cold accretion or recycled
material falling back onto galaxies, yet they are the most direct
measure of accreting gas to date.

Using quasar sight-lines, we find that the distribution of velocity
separations between {\MgII} absorption and their host galaxies tends
to be Gaussian with a mean offset of 16~{\kms} and dispersion of about
140~{\kms} \citep{chen10}. Although there are much higher velocity
extremes, typically expected for outflows, it is interesting that the
velocity range is more typical of galaxy rotation speeds having
masses close-to or less-than that of the Milky Way. 

\begin{figure}[t]
\includegraphics[scale=0.46]{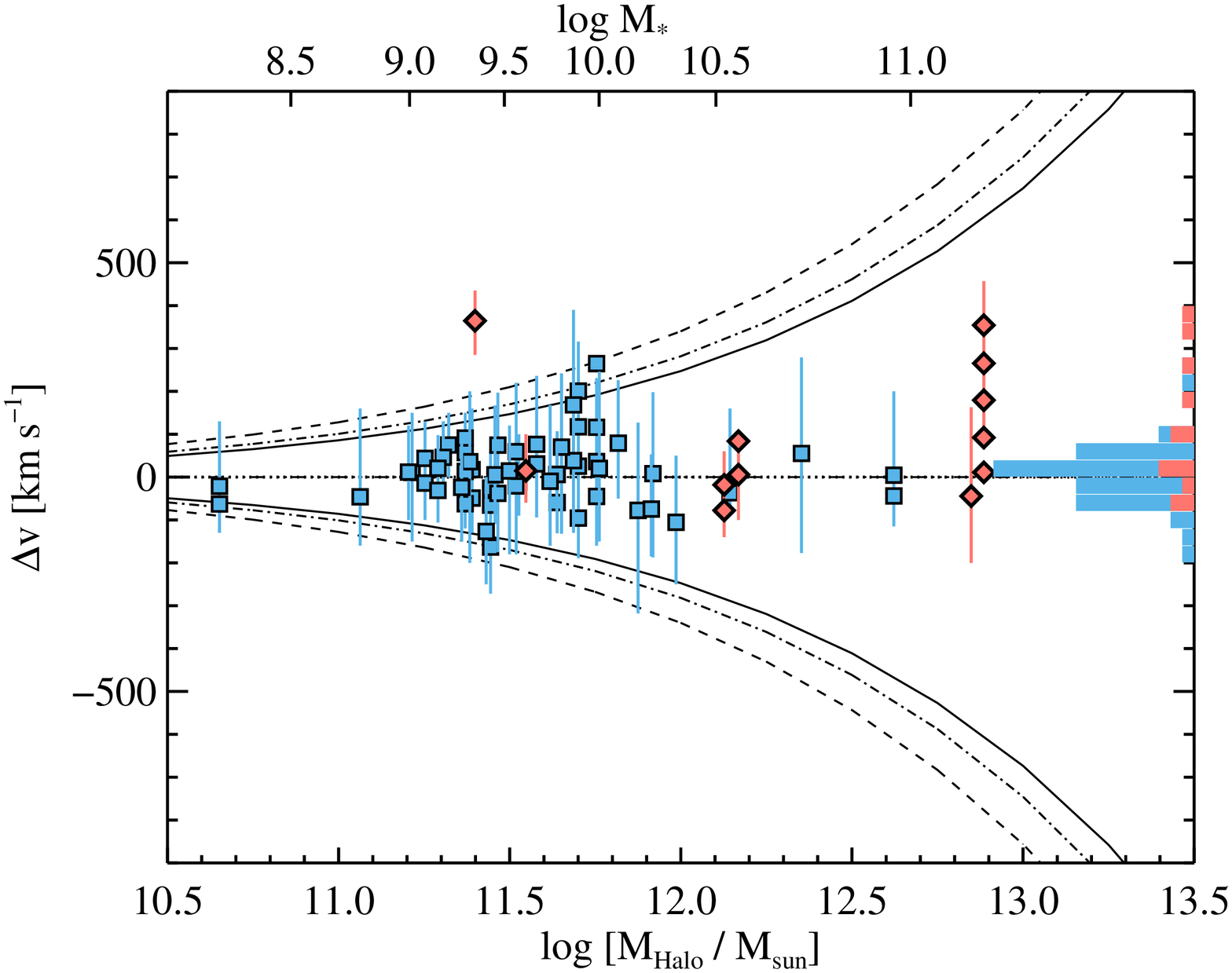}
%
%
\caption{Combined {\CIV} \citep{bordoloi14a} and {\OVI}
  \citep{tumlinson11} absorption velocity centroids with respect to
  the systemic redshift of their host galaxies as a function of the
  inferred dark matter halo mass for star-forming (blue squares) and
  passive (red diamond) galaxies. The range bars indicate the maximum
  projected kinematic extent of each absorption system. The histogram
  represents the distribution of individual component velocities. The
  dashed lines show the mass-dependent escape velocities at R = 50,
  100, and 150 kpc, respectively. Note that all absorption-line
  systems appear to be bound to their halos and have velocities (and
  velocity ranges) comparable to galaxy circular velocities. This
  means that both outflowing and accreting gas could give rise to the
  observed kinematics. Image courtesy of Jason Tumlinson, Rongmon
  Bordoloi and the COS$-$Halos team.}
\label{fig:escape}       
\end{figure}

When galaxy masses are known, one can compare the relative galaxy and
absorption velocities to those of the escape velocities of their
halos. Figure~\ref{fig:escape} shows escape velocities, computed for
spherically symmetric Navarro--Frenk--White dark matter halo profile
\citep{nfw96}, as a function of halo mass. It can be seen that very
few absorption velocity centroids exceed the estimated galaxy halo
escape velocities. This is surprisingly true for a full range of
galaxy masses from dwarf galaxies probed by {\CIV} absorption
\citep{bordoloi14a} to more massive galaxies probed by {\OVI}
\citep{tumlinson11}. Consistently, it has been found that the vast
majority of absorption systems that reside within one galaxy virial
radius are bound to the dark matter halos of their hosts
\citep{stocke13,mathes14,tumlinson11,bordoloi14a,ho16}.  It is worth
noting that some of the velocity ranges covered by the absorption
profiles are comparable to the escape velocities but are typically for
the far wings of the profiles where the gas column densities are the
lowest.  Therefore two possible scenarios, or a combination thereof,
can be drawn: 1) gas that is traced by absorption can be driven into
the halo by star formation driven outflows and eventually fall back
onto the galaxy (known as recycled winds) and/or 2) the gas is new
material accreting from the intergalactic medium. With these data
alone, we likely cannot distinguish between these scenarios.

To test how the circumgalactic gas is kinematically coupled to their
galaxy hosts, \citet{steidel02} presented the first rotation curves of
five intermediate-redshift {\MgII} selected absorbing
galaxies. Interestingly, they found that for four of the five cases,
the absorption velocities lie entirely to one side of the galaxy
systemic redshift and consistent with the side expected for rotation.
Using simple thick disk-halo models, they concluded that the bulk of
{\MgII} gas velocities could be explained by an extension of disk
rotation with some velocity lag \citep{steidel02}. This was further
confirmed by \citet{kacprzak10a} who also showed that infalling gas or
lagging rotation is required to explain the gas kinematics. Using
cosmological hydrodynamical galaxy simulations to replicate their data
allowed them to concluding that coherently rotating accreting gas is
likely responsible for the observed kinematic offset.

There have now been over 50 galaxies/absorbers pairs that have been
compared this way and the vast majority exhibit disk-like and/or
accretion kinematics
\citep{steidel02,chen05,kacprzak10a,kacprzak11a,bouche13,burchett13,keeney13,jorgenson14,diamond-stanic16,bouche16,ho16}
and some show outflowing wind signatures
\citep{ellsion03,kacprzak10a,bouche12,schroetter15,muzahid16,schroetter16}
while some exhibiting group dynamics
\citep{lehner09,kacprzak10b,bielby16,peroux16}.  Even for systems with
multiple quasar sight-lines \citep{bowen16} or for multiply-lensed
quasars near known foreground galaxies \citep{chen14} provide the same
kinematic evidence that a co-rotating disk with either some lagging
rotation or accretion is required to reproduce the observed absorption
kinematics.  One caveat is that above works have a range of galaxy
inclinations and quasar sight-line azimuthal angles, which could
complicate the conclusions drawn. It is likely best to select galaxies
where the quasar is located along the projected major axis where
accreted gas is expected to be located.

 \begin{figure}[b]
\includegraphics[scale=0.27]{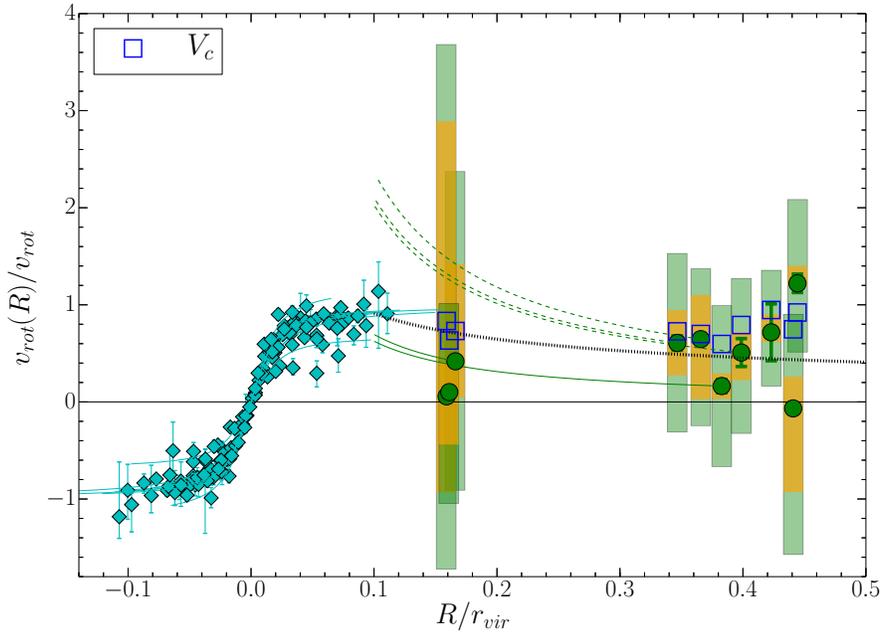}
%
%
\caption{Galaxy rotation curves of 10 galaxies, normalized by the
  rotation speed as a fraction of the halo virial radius, are compared
  to the kinematics of their circumgalactic gas. The {\MgII}
  absorption velocities are deprojected such that the velocity shown
  represents the tangential motion in the disk plane that would give
  to the observed sight-line velocities.  The measured and intrinsic
  velocity range of each {\MgII} absorption system after deprojection
  is indicated by the green and orange bars with the {\MgII} absorption 
velocity along the quasar sight-line (green circles). Note that the absorption systems align with the expected 
  side for extended disk rotation. Also shown are dark matter halo rotation speed models (blue
  squares) and the Keplerian fall off from the galaxy rotation curves, which sets a 
  lower limits on the rotation speed in the circumgalactic
  medium (dashed, black line). The cyan curves illustrate constant
  $Rv_{rot}(R)$ and indicates that the infalling gas would have specific angular
  momentum at least as large as that in the galactic disk.}
\label{fig:rot}       
\end{figure}

\citet{ho16} designed an experiment where they selected {\MgII}
absorbers associated with highly inclined ($i>43$~degrees)
star-forming galaxies with quasars sight-lines passing within
$30$~degrees of the projected major axis.  Presented in
Figure~\ref{fig:rot} are the rotation curves and the velocity spread
of the absorption shown as a function of distance within the galaxy
viral radius.  It is clear that there strong correlation between the
{\MgII} absorption velocities and the galaxy rotation velocities.  The
majority of the {\MgII} equivalent widths are detected at velocities
less than the actual rotation speed of the dark matter halo (blue
squares), while the Keplerian fall-off from the measured rotation
curve provides lower limits on the rotation speed of the
circumgalactic medium (dashed, black line). The cyan curves illustrate
constant $Rv_{rot}(R)$ and show that the infalling gas would have
specific angular momentum at least as large as that in the galactic
disk, for which some of the gas has comparable specific angular
momentum. The {\MgII} absorption-line velocity widths cannot be
generated with circular disk-like orbit and a simple disk model with a
radial inflowing accretion reproduced the data quite well
\citep{ho16}.

We have shown that lagging or infalling gas appears to be a common
kinematic signature of the circumgalactic gas near star-forming
galaxies from 0.1$\leq z \leq 2.5$ and is consistently seen for a
range of galaxy inclination and position angles. These observations
are consistent with current simulations that show that large
co-rotating gaseous structures in the halo of the galaxy that are
fueled, aligned, and kinematically connected to filamentary gas infall
along the cosmic
web. \citep{stewart11,danovich12,stewart13,danovich15,stewart16}. The
predictions and results from simulations are discussed further in
the chapters by Kyle Stewart and Claude-Andr\'e Fauch\`er-Giguere.
\citet{stewart16} demonstrated that there is a
qualitative agreement among the majority of cosmological simulations
and that the buildup of high angular momentum halo gas and the
formation of cold flow disks are likely a robust prediction of
$\Lambda$CDM. These simulations naturally predict that accreted gas
can be observationally distinguishable from outflowing gas from its
kinematic signature of large one-sided velocity offsets. Thus, it is
plausible we have already observed gas accretion through these
kinematic velocity offsets mentioned above.

It is also important to note that the vast majority of systems
discussed above have column densities typical of Lyman Limit Systems
(N(\HI)$>10^{17.2}$~cm$^{-2}$), which are exclusively associated with
galaxies. At the lowest column densities of
N(\HI)$<10^{14}$~cm$^{-2}$, {\Lya} absorption was found to not mimic
the rotation of, and/or accretion onto, galaxies derived from {\HI}
observations \citep{cote05}. Maybe this is not unexpected since this
low column density gas is not likely to be associated with galaxies
and likely associated with the {\Lya} forest \citep[gas associated with
galaxies have N(\HI)$>10^{14.5}$~cm$^{-2}-$][]{rudie12}.

\section{Circumgalactic and Galaxy Gas-Phase Metallicities}
\label{sec:3}

It is possible that metallicity can be used to determine the origins
of absorbing gas observed around galaxies since outflows are expected
to be metal-enriched while accreted gas should have lower metallicity
\citep[e.g.,][]{shen12}.  Gas accretion is expected to be metal poor
but not purely pristine given that the first generations of
Population~III stars have likely enriched the gas to
10$^{-4}$~$Z_\odot$ by a redshift of $15-20$
\citep[e.g.,][]{yoshida04}.  In fact, there is a metallicity floor
whereby it is rare to find absorption systems with metallicities much
lower than 10$^{-3}$~$Z_\odot$ even out to $z\sim5$
\citep{prochaska03,penprase10,cooke11,battisti12,rafelski12,jorgenson13,cooke15,cooper15,fumagalli16a,lehner16,quiret16}. Although
a few systems do have metallicities $<$10$^{-3}$~$Z_\odot$
\citep{fumagalli11a,cooke11,cooke15,lehner16} and possibly have
Population~III abundance patterns \citep{crighton16}.  Cosmological
simulations predict gas accretion metallicities should be between
10$^{-3}-10^{-0.5}$~~$Z_\odot$, which is dependent on redshift and
halo mass
\citep{fumagalli11b,oppenheimer12,vandevoort12,shen13,kacprzak16},
however this metallicity range does have some overlap with the
metallicities of recycled outflowing gas.

There has been an abundance of studies that have identified galaxies
with circumgalactic gas metallicity measurements in an effort to
determine the source of the absorption.  The general census shows that
absorption systems near galaxies are either metal-poor with
metallicities between $-2<$[X/H]$<-1$
\citep{tripp05,cooksey08,kacprzak10b,ribaudo11,thom11,churchill12,bouche13,crighton13,stocke13,kacprzak14,crighton15,muzahid15,bouche16,fumagalli16b,rahmani16}
or metal-enriched with metallicities of [X/H]$>-0.7$
\citep{chen05,lehner09,peroux11,bregman13,krogager13,meiring13,stocke13,crighton15,muzahid15,muzahid16,peroux16,rahmani16}. Determining
the fraction of metal-rich and metal-poor systems is complicated since
all the aforementioned studies have a range of observational biases
due to the way the targets were selected -- some selected from the
presence of metal-lines.  A clear complication of tracing the
circumgalactic gas using metal-lines as tracers may bias you towards
metal enriched systems. Ideally, the best way to avoid such a biases
is to select absorption systems by hydrogen only.

Selecting only by hydrogen (but not necessarily with known galaxy
hosts), has shown that the metallicity distribution of all Lyman limit
systems below $z<1$ appear to have a bi-modal distribution
\citep{lehner13,wotta16}. The shape of Lyman limit systems
($17.2 < $log N({\HI}) $< 17.7$ in their study) metallicity bimodality
distribution could be explained by outflows producing the high
metallicity peak ([X/H]$\sim -0.3$) while accreting/recycled gas could
produce the low metallicity peak ([X/H] $\sim -1.9$)
\citep{wotta16}. Interestingly, between $2<z<3.5$, the metallicity
distribution of Lyman limit systems uni-modal peaking at [X/H]$=-2$,
in contrast to the bimodal distribution seen at $z<1$ \citep{fumagalli11a,lehner16}. Therefore it is
likely that there exists a vast reservoir of metal-poor cool gas that
can accrete onto galaxies at high redshift and outflows build up the
circumgalactic medium at a later time. These results
are discussed in detail in the chapter by Nicolas Lehner.

 \begin{figure}[b]
\includegraphics[scale=0.42]{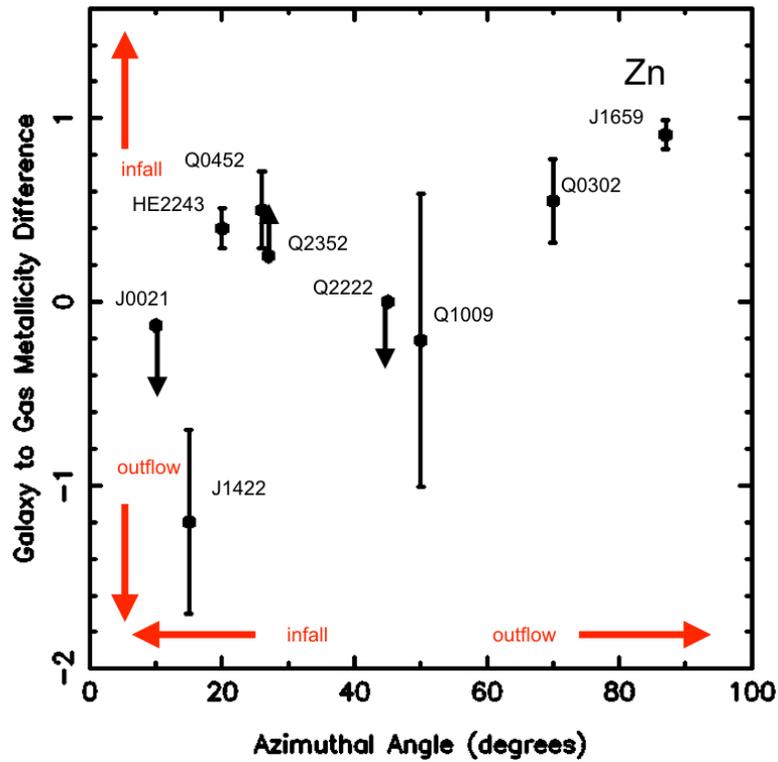}
%
%
\caption{Metallicity difference between the host galaxy and absorber
  as a function of azimuthal angle. In this plot, accreted gas is
  expected to reside in the upper-left corner (for low/co-planer
  azimuthal angle and high metallicity difference), while outflowing
  gas should reside in the lower-right corner (high/minor axis
  azimuthal angle and metallicity similar to the host galaxy) as
  indicated by the red arrows. Note outflowing gas appears to be
  metal-poor, while accreting gas exhibits a range in
  metallicity. Additional observations are necessary to better relate
  metallicity and geometry in gas flows as only a minor
  anti-correlation is currently measured. Image from
  \citet{peroux16}.}
\label{fig:azxh}       
\end{figure}

The overall knowledge of the metallicity distribution of the
circumgalactic medium provides critical clues to the physics of gas
cycles of galaxies. The bi-modal metallicity distribution is
suggestive that we are observing both outflows and accretion, but our
assumptions rely strongly on predictions from simulations, which still
have some issues with modeling the circumgalactic medium since it is
extremely feedback dependent. One way to determine if we are observing
outflows and gas accretion is to combine our expectation of the gas
geometry of accretion being along the minor axis and metal poor, while
outflows are metal-enrich and ejected along the minor axis.

Preliminary work by \citet{peroux16}, using nine galaxies, indicate
that there is a very weak anti-correlation with metallicity and
azimuthal angle.  Figure~\ref{fig:azxh} show the relative metallicity
of the absorbing gas with respect to the host galaxy metallicity as a
function of azimuthal angle, which are two independent indicators of
gas flow origins.  Note that in the figure a positive difference in
metallicity indicates a circumgalactic medium metallicity lower than
that of the host galaxy's HII regions (expected for metal-poor
accreting gas) while a negative value indicates a higher
circumgalactic medium metallicity than the host galaxy (expected for
metal-enriched outflows). For the few objects shown in
Figure~\ref{fig:azxh}, there is not clear correlation as expected
under simple geometric and metallicity assumptions. Note that,
different from expectations, there does not appear to be any high
metallicity gas at high azimuthal angles. At low low azimuthal angles,
there are a range of metallicity differences including negative
values, which is unexpected for accreting gas. Given the few number of
systems and only a weak anti-correlation, more systems are required to
understand if there is a relation between the spatial location of the
circumgalactic medium and metallicity.  This is an active area of
research and may be the most promising avenue to peruse in the future.

\section{Putting it all Together}
\label{sec:4}

The previous sections describe the individual geometric, kinematic and
metallicity indicators as evidence for cold-mode accretion. On their
own, they are quite suggestive that we have detected signatures of gas
accretion, however, combining all of these accretion indicators
together can provide quite compelling evidence.

There are a few of such examples that exist where some point to their
circumgalactic medium originating from metal enriched outflows along
the galaxy minor axis \citep{kacprzak14,muzahid15} or the
circumgalactic medium is kinematically consistent with gas arising
from tidal/streams or interacting galaxy groups
\citep{kacprzak11b,muzahid16,peroux16} or even enriched gas that is
being recycled along the galaxy major axis \citep{bouche16}.

\citet{bouche13} presented a nice example of a moderately inclined
$z=2.3$ star-forming galaxy where the quasar sight-line is within
20~degrees of the galaxy's projected major axis. They derived the
galaxy rotation field using IFU observations and found that the
kinematics of circumgalactic medium at 26~kpc away could be reproduced
by a combination of an extended rotating disk and radial gas
accretion.  The metallicity of the circumgalactic medium is $-0.72$,
which is typically for gas accretion metallicities from cosmological
simulations at these redshifts \citep{vandevoort12,kacprzak16}. The
mass inflow rate was estimated to be between
$30-60$~M$_{\odot}$yr$^{-1}$, which is similar to the galaxy
star-formation rate of 33~~M$_{\odot}$yr$^{-1}$ and suggestive that
there is a balance between gas accretion and star-formation activity.
This particular system is described in detail in the chapter by
Nicolas Bouch\'e.

In a slightly different case, a star-forming galaxy at $z=0.66$ was
examined where the quasar sightline is within 3~degrees of the minor
axis at a distance of 104~kpc \citep{kacprzak12b}. Contrary to
expectations, they identify a cool gas phase with metallicity
$\sim-1.7$ that has kinematics consistent with a accretion or lagging
halo model. Furthermore, they also identify a warm collisionally
ionized phase that also has low metallicity ($\sim-2.2$). The warm gas
phase is kinematically consistent with both radial outflows or radial
accretion. Given the metallicities and kinematics, they conclude that
the gas is accreting onto the galaxy, however this is contrary to the
previously discussed interpretations that absorption found along a the
projected minor axis is typically associated outflows. This could be a
case where there is a miss-alignment with the accreting filaments and
the disk or is an example of the three filaments typically predicted
by simulations \citep[e.g.,][]{dekel09,danovich12}, thus making it
unlikely that all accreting gas is co-planer.

Detailed studies like these are one of the best ways of constraining
the origins of the circumgalactic medium and to help us understand how
gas accretion works. Larger samples are required to build up a
statistical sample of systems to negate cosmic variance. Large IFU
instruments like the Multi Unit Spectroscopic Explorer (MUSE) will
provide ample full field imaging/spectra datasets and help us build up
these larger samples with less observational time. Gas accretion
studies using IFUs are further discussed in the chapter by 
Nicolas Bouch\'e.

 \begin{figure}
\begin{center}
\includegraphics[scale=0.29]{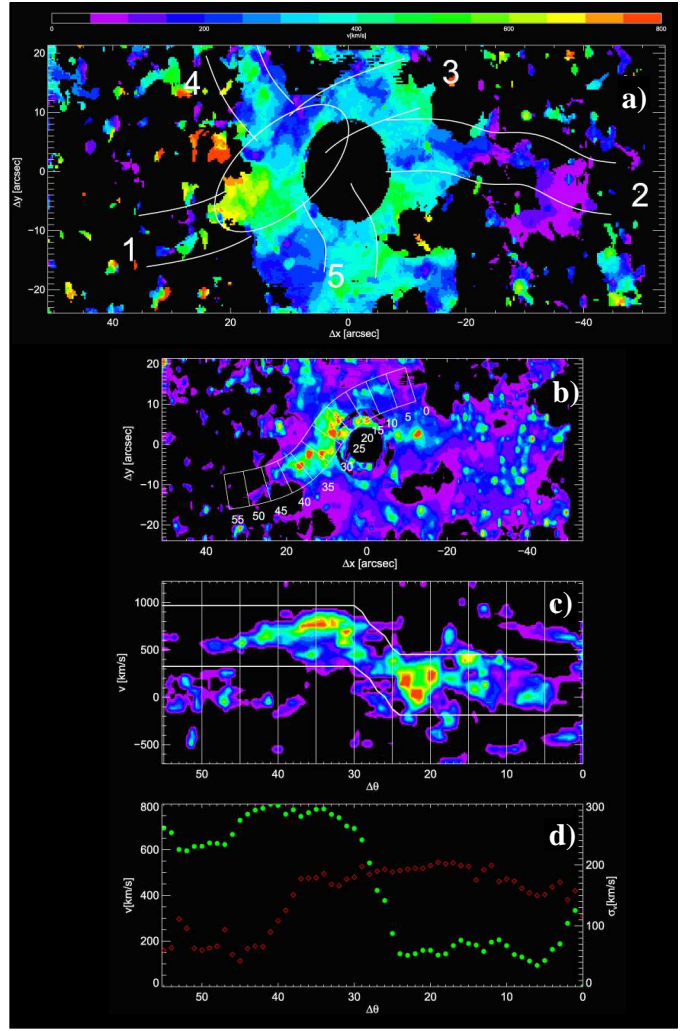}
\end{center}
%
%
\caption{ a) Mean velocity obtained with an intensity-weighted
  velocity moment within a narrow 7~{\AA} band image shown in panel
  b). The data were obtained using Palomar cosmic web imager. The disk
  and filamentary candidates are indicated. -- b) A pseudo slit is
  placed over the narrow-band image covering the disk and filaments 1
  and 3 indicated by the curved white lines with tick marks indicating
  the distance in arcseconds along the slit. -- c) The narrow-band
  image produced using the sheared velocity window slit indicating the
  distance along the slit as in panel b) -- d) The mean velocity
  (green) and velocity dispersion (red) from the above panels. Note
  observed rotation with high velocity dispersions seen along the
  filament. This is also seen for the remaining filaments (not shown
  here). This is potentially one of the first direct images of cold gas
  accretion. Image courtesy of Chris Martin and modified from
  \citet{martin16}.}
\label{fig:martin}       
\end{figure}

\section{Direct Imaging of Gas Accretion}
\label{sec:4A}

All of the aforementioned efforts are possibly direct observations of
gas accretion, yet the scientific community is not satisfied since it
is difficult to prove conclusively that we have detected cold gas
accretion onto galaxies.  The only way to conclusively do so is by
direct spectral imaging of cosmic flows onto galaxies. Obviously this
is quite difficult due to the faintness and low gas column densities
of these cold-flow filaments. However, we may be on the verge of being
able to detect these cold gas-flows using the new generation of ultra
sensitive instruments such as the Keck Comic Web Imager.

It is also possible that we have already observed comic accretion onto
two separate quasar hosts \citep{martin15,martin16} with one example
shown in Figure~\ref{fig:martin}. The figure shows a narrow-band image
obtained with the Palomar Comic Web Imager. The extended nebula and
filaments are likely illuminated by the nearby quasar. The streams, as
indicted in the figure, extend out to $\sim 160-230$~kpc. It was
determined that this nebula/disk and filaments are well fit to a
rotating disk model with are dark matter halo mass of
$\sim$10$^{12.5}$~M$_{\odot}$ with a circular velocity of
$\sim$350~{\kms} at the viral radius of 125~kpc \citep{martin16}. They
further found that by adding gas accretion with velocities of
$80-100$~{\kms} improved their kinematic model fit. Furthermore, they
estimate the baryonic spin parameter is 3 times higher than that of
the dark matter halo and has an orbital period of 1.9~Gyr. The
high-angular momentum and the well fitted inflowing stable disk is
consistent with the predictions of cold accretion from cosmological
simulations
\citep{stewart11,danovich12,stewart13,danovich15,stewart16}.

It is possible that we have direct evidence of cold accretion already,
but it has yet to be observed and quantified for typical star-forming
galaxies (which is much harder). These types of ultra-sensitive
instruments will potentially allow us to directly image cosmic
accretion in the near future.

\section{Summary}
\label{sec:5}

Although we only have candidate detections of gas accretion, the
community has provided a body of evidence indicating that cosmic
accretion exists and is occurring over a large range of
redshifts. Individual studies on their own are only suggestive that
this accretion is occurring, but taking all the aforementioned aspects
together does paint a nice picture of low metallicity gas accreting
along co-planar co-rotating disk/filaments. The picture seems nice and
simple but understanding how previously ejected gas is also recycled
back onto the galaxy is also important since it can possibly mimic gas
accretion signatures \citep{oppenheimer12,ford14}.

Simulations predict the cross-section of cold flows to be as low as
5\% \citep{faucher-Giguere11,fumagalli11a,kimm11,goerdt12} and that
the metal-poor cold flow streams signatures should be overwhelmed by
the metal-rich outflows signatures detected in absorption
spectra. However, signatures of intergalactic cold gas accretion seem
to be quite frequent. Thus, either the simulations are under-producing
gas accretion cross-sections due to various reasons such as dust,
resolution effects, self-shielding and/or magnetic fields, or the
observations are selecting more than just outflowing and accreting
gas. Trying to address how recycled winds fit into the picture, and
understanding how we can differentiate them from gas accretion, will
possibly aide in resolving this issue.  It is great to see however,
how the simulations and observations have been working together in our
community to try and understand the gas cycles of galaxies.

Aside from observing accretion directly, the way forward now is to
combine geometry, kinematics and metallicity at a range of epochs to
try to understand how gas accretion occurs. Hopefully in the near
future, we will be able to directly image gas accretion, putting to
rest one of the most debated issues in the circumgalactic field.

\begin{acknowledgement}
  GGK acknowledges the support of the Australian Research Council
  through the award of a Future Fellowship (FT140100933). Thanks to
  Nikole Nielsen for providing comments, and to Nikole Nielsen,
  Stephanie Ho and Rongmon Bordoloi for providing additional new plots
  for this Chapter.
\end{acknowledgement}
%

\input{referenc-kacprzak}
\end{document}

%% file: referenc-kacprzak.tex
\def\aj{{AJ}}                   
\def\araa{{ARA\&A}}             
\def\apj{{ApJ}}                 
\def\apjl{{ApJ}}                
\def\apjs{{ApJS}}               
\def\ao{{Appl.~Opt.}}           
\def\apss{{Ap\&SS}}             
\def\aap{{A\&A}}                
\def\aapr{{A\&A~Rev.}}          
\def\aaps{{A\&AS}}              
\def\azh{{AZh}}                 
\def\baas{{BAAS}}               
\def\jrasc{{JRASC}}             
\def\memras{{MmRAS}}            
\def\mnras{{MNRAS}}             
\def\pra{{Phys.~Rev.~A}}        
\def\prb{{Phys.~Rev.~B}}        
\def\prc{{Phys.~Rev.~C}}        
\def\prd{{Phys.~Rev.~D}}        
\def\pre{{Phys.~Rev.~E}}        
\def\prl{{Phys.~Rev.~Lett.}}    
\def\pasa{{PASA}}
\def\pasp{{PASP}}
\def\pasj{{PASJ}}               
\def\qjras{{QJRAS}}             
\def\skytel{{S\&T}}             
\def\solphys{{Sol.~Phys.}}      
\def\sovast{{Soviet~Ast.}}      
\def\ssr{{Space~Sci.~Rev.}}     
\def\zap{{ZAp}}                 
\def\nat{{Nature}}              
\def\iaucirc{{IAU~Circ.}}
\def\aplett{{Astrophys.~Lett.}}
\def\apspr{{Astrophys.~Space~Phys.~Res.}}
\def\bain{{Bull.~Astron.~Inst.~Netherlands}}
\def\fcp{{Fund.~Cosmic~Phys.}}
\def\gca{{Geochim.~Cosmochim.~Acta}}
\def\grl{{Geophys.~Res.~Lett.}}
\def\jcp{{J.~Chem.~Phys.}}      
\def\jgr{{J.~Geophys.~Res.}}    
\def\jqsrt{{J.~Quant.~Spec.~Radiat.~Transf.}}
\def\memsai{{Mem.~Soc.~Astron.~Italiana}}
\def\nphysa{{Nucl.~Phys.~A}}
\def\physrep{{Phys.~Rep.}}
\def\physscr{{Phys.~Scr}}
\def\planss{{Planet.~Space~Sci.}}
\def\procspie{{Proc.~SPIE}}

%% file: kacprzak.bbl
\begin{thebibliography}{99.}%
%


\bibitem[Adelberger et al.(2005)]{adelberger05} Adelberger, K.~L., Shapley, A.~E., Steidel, C.~C., et al.\ 2005, \apj, 629, 636 
\bibitem[Battisti et al.(2012)]{battisti12} Battisti, A.~J., Meiring, J.~D., Tripp, T.~M., et al.\ 2012, \apj, 744, 93 
\bibitem[Bergeron(1986)]{bergeron86} Bergeron, J.\ 1986, \aap, 155, L8 
\bibitem[Bielby et al.(2016)]{bielby16} Bielby, R., Crighton, N.~H.~M., Fumagalli, M., et al.\ 2016, arXiv:1607.03386 
\bibitem[Birnboim \& Dekel(2003)]{birnboim03} Birnboim, Y., \& Dekel, A.\ 2003, \mnras, 345, 349 
\bibitem[Boksenberg \& Sargent (1978)]{boksenberg78} Boksenberg, A., \& Sargent, W.~L.~W.\ 1978, \apj, 220, 42 
\bibitem[Bordoloi et al.(2011)]{bordoloi11} Bordoloi, R., Lilly, S.~J., Knobel, C., et al.\ 2011, \apj, 743, 10 
\bibitem[Bordoloi et al.(2014b)]{bordoloi14b} Bordoloi, R., Lilly, S.~J., Kacprzak, G.~G., \& Churchill, C.~W.\ 2014, \apj, 784, 108 
\bibitem[Bordoloi et al.(2014a)]{bordoloi14a} Bordoloi, R., Tumlinson, J., Werk, J.~K., et al.\ 2014, \apj, 796, 136 
\bibitem[Borthakur et al.(2015)]{borthakur15} Borthakur, S., Heckman, T., Tumlinson, J., et al.\ 2015, \apj, 813, 46 
\bibitem[Borthakur et al.(2016)]{borthakur16} Borthakur, S., Heckman, T., Tumlinson, J., et al.\ 2016, arXiv:1609.06308 
\bibitem[Bouch{\'e} et al.(2016)]{bouche16} Bouch{\'e}, N., Finley, H., Schroetter, I., et al.\ 2016, \apj, 820, 121 
\bibitem[Bouch{\'e} et al.(2012)]{bouche12} Bouch{\'e}, N., Hohensee, W., Vargas, R., et al.\ 2012, \mnras, 426, 801 
\bibitem[Bouch{\'e} et al.(2013)]{bouche13} Bouch{\'e}, N., Murphy, M.~T., Kacprzak, G.~G., et al.\ 2013, Science, 341, 50 
\bibitem[Bowen et al.(2016)]{bowen16} Bowen, D.~V., Chelouche, D., Jenkins, E.~B., et al.\ 2016, \apj, 826, 50 
\bibitem[Bregman et al.(2013)]{bregman13} Bregman, J.~N., Miller, E.~D., Seitzer, P., Cowley, C.~R., \& Miller, M.~J.\ 2013, \apj, 766, 57 
\bibitem[Brooks et al.(2009)]{brooks09} Brooks, A.~M., Governato, F., Quinn, T., Brook, C.~B., \& Wadsley, J.\ 2009, \apj, 694, 396
\bibitem[Burchett et al.(2015)]{burchett15} Burchett, J.~N., Tripp, T.~M., Prochaska, J.~X., et al.\ 2015, \apj, 815, 91 
\bibitem[Burchett et al.(2013)]{burchett13} Burchett, J.~N., Tripp, T.~M., Werk, J.~K., et al.\ 2013, \apjl, 779, L17 
\bibitem[Charlton \& Churchill(1996)]{charlton96} Charlton, J.~C., \& Churchill, C.~W.\ 1996, \apj, 465, 631 
\bibitem[Charlton \& Churchill(1998)]{charlton98} Charlton, J.~C., \& Churchill, C.~W.\ 1998, \apj, 499, 181 
\bibitem[Chen et al.(2005)]{chen05} Chen, H.-W., Kennicutt, R.~C., Jr., \& Rauch, M.\ 2005, \apj, 620, 703 
\bibitem[Churchill et al.(2012)]{churchill12} Churchill, C.~W., Kacprzak, G.~G., Steidel, C.~C., et al.\ 2012, \apj, 760, 68 
\bibitem[Churchill et al.(2013a)]{churchill13a} Churchill, C.~W., Nielsen, N.~M., Kacprzak, G.~G., \& Trujillo-Gomez, S.\ 2013, \apjl, 763, L42 
\bibitem[Churchill et al.(2013b)]{churchill13b} Churchill, C.~W., Trujillo-Gomez, S., Nielsen, N.~M., \& Kacprzak, G.~G.\ 2013, \apj, 779, 87 
\bibitem[Chen et al.(2010)]{chen10} Chen, H.-W., Helsby, J.~E., Gauthier, J.-R., et al.\ 2010, \apj, 714, 1521 
\bibitem[Chen et al.(2014)]{chen14} Chen, H.-W., Gauthier, J.-R., Sharon, K., et al.\ 2014, \mnras, 438, 1435 
\bibitem[Chen et al.(2005)]{chen05} Chen, H.-W., Kennicutt, R.~C., Jr., \& Rauch, M.\ 2005, \apj, 620, 703 
\bibitem[Chen et al.(2001a)]{chen01a} Chen, H.-W., Lanzetta, K.~M., \& Webb, J.~K.\ 2001, \apj, 556, 158 
\bibitem[Chen et al.(2001b)]{chen01b} Chen, H.-W., Lanzetta, K.~M., Webb, J.~K., \& Barcons, X.\ 2001, \apj, 559, 654 
\bibitem[Chen \& Mulchaey(2009)]{chen09} Chen, H.-W., \& Mulchaey, J.~S.\ 2009, \apj, 701, 1219 
\bibitem[C{\^o}t{\'e} et al.(2005)]{cote05} C{\^o}t{\'e}, S., Wyse, R.~F.~G., Carignan, C., Freeman, K.~C., \& Broadhurst, T.\ 2005, \apj, 618, 178 
\bibitem[Cooke et al.(2015)]{cooke15} Cooke, R.~J., Pettini, M., \& Jorgenson, R.~A.\ 2015, \apj, 800, 12 
\bibitem[Cooke et al.(2011)]{cooke11} Cooke, R., Pettini, M., Steidel, C.~C., Rudie, G.~C., \& Nissen, P.~E.\ 2011, \mnras, 417, 1534 
\bibitem[Cooksey et al.(2008)]{cooksey08} Cooksey, K.~L., Prochaska, J.~X., Chen, H.-W., Mulchaey, J.~S., \& Weiner, B.~J.\ 2008, \apj, 676, 262 
\bibitem[Cooper et al.(2015)]{cooper15} Cooper, T.~J., Simcoe, R.~A., Cooksey, K.~L., O'Meara, J.~M., \& Torrey, P.\ 2015, \apj, 812, 58 
\bibitem[Crighton et al.(2013)]{crighton13} Crighton, N.~H.~M., Hennawi, J.~F., \& Prochaska, J.~X.\ 2013, \apjl, 776, L18 
\bibitem[Crighton et al.(2015)]{crighton15} Crighton, N.~H.~M., Hennawi, J.~F., Simcoe, R.~A., et al.\ 2015, \mnras, 446, 18
\bibitem[Crighton et al.(2016)]{crighton16} Crighton, N.~H.~M., O'Meara, J.~M., \& Murphy, M.~T.\ 2016, \mnras, 457, L44 
\bibitem[Danforth \& Shull(2008)]{danforth08} Danforth, C.~W., \& Shull, J.~M.\ 2008, \apj, 679, 194
\bibitem[Danovich et al.(2012)]{danovich12} Danovich, M., Dekel, A., Hahn, O., \& Teyssier, R.\ 2012, \mnras, 422, 1732 
\bibitem[Danovich et al.(2015)]{danovich15} Danovich, M., Dekel, A., Hahn, O., Ceverino, D., \& Primack, J.\ 2015, \mnras, 449, 2087 
\bibitem[Dekel \& Birnboim(2006)]{dekel06} Dekel, A., \& Birnboim, Y.\ 2006, \mnras, 368, 2 
\bibitem[Dekel et al.(2009)]{dekel09} Dekel, A., Birnboim, Y., Engel, G., et al.\ 2009, \nat, 457, 451
\bibitem[Diamond-Stanic et al.(2016)]{diamond-stanic16} Diamond-Stanic, A.~M., Coil, A.~L., Moustakas, J., et al.\ 2016, \apj, 824, 24 
\bibitem[Dutta et al.(2016)]{dutta16} Dutta, R., Srianand, R., Gupta, N., et al.\ 2016, arXiv:1610.05316 
\bibitem[Ellison et al.(2003)]{ellsion03} Ellison, S.~L., Mall{\'e}n-Ornelas, G., \& Sawicki, M.\ 2003, \apj, 589, 709 
\bibitem[Faucher-Gigu{\`e}re \& Kere{\v s}(2011)]{faucher-Giguere11} Faucher-Gigu{\`e}re, C.-A., \& Kere{\v s}, D.\ 2011, \mnras, 412, L118 
\bibitem[Ford et al.(2014)]{ford14} Ford, A.~B., Dav{\'e}, R., Oppenheimer, B.~D., et al.\ 2014, \mnras, 444, 1260 
\bibitem[Fumagalli et al.(2016b)]{fumagalli16b} Fumagalli, M., Cantalupo, S., Dekel, A., et al.\ 2016, \mnras, 462, 1978 
\bibitem[Fumagalli et al.(2011b)]{fumagalli11a} Fumagalli, M., Prochaska, J.~X., Kasen, D., et al.\ 2011, \mnras, 418, 1796 
\bibitem[Fumagalli et al.(2011a)]{fumagalli11b} Fumagalli, M., O'Meara, J.~M., \& Prochaska, J.~X.\ 2011, Science, 334, 1245 
\bibitem[Fumagalli et al.(2016a)]{fumagalli16a} Fumagalli, M., O'Meara, J.~M., \& Prochaska, J.~X.\ 2016, \mnras, 455, 4100 
\bibitem[Goerdt et al.(2012)]{goerdt12} Goerdt, T., Dekel, A., Sternberg, A., Gnat, O., \& Ceverino, D.\ 2012, \mnras, 424, 2292 
\bibitem[Guillemin \& Bergeron(1997)]{gb97} Guillemin, P., \& Bergeron, J. 1997, A\&A, 328, 499
\bibitem[Ho et al.(2016)]{ho16} Ho, S.~H., Martin, C.~L., Kacprzak G. G., Churchull, C.~W. 2016, \apj, submitted
\bibitem[Johnson et al.(2015)]{johnson15} Johnson, S.~D., Chen, H.-W., \& Mulchaey, J.~S.\ 2015, \mnras, 449, 3263
\bibitem[Johnson et al.(2013)]{johnson13} Johnson, S.~D., Chen, H.-W., \& Mulchaey, J.~S.\ 2013, \mnras, 434, 1765
\bibitem[Jorgenson et al.(2013)]{jorgenson13} Jorgenson, R.~A., Murphy, M.~T., \& Thompson, R.\ 2013, \mnras, 435, 482 
\bibitem[Jorgenson \& Wolfe(2014)]{jorgenson14} Jorgenson, R.~A., \& Wolfe, A.~M.\ 2014, \apj, 785, 16 
\bibitem[Kacprzak et al.(2011a)]{kacprzak11a} Kacprzak, G.~G., Churchill, C.~W., Barton, E.~J., \& Cooke, J.\ 2011, \apj, 733, 105 
\bibitem[Kacprzak et al.(2010a)]{kacprzak10a} Kacprzak, G.~G., Churchill, C.~W., Ceverino, D., et al. 2010, ApJ, 711, 533
\bibitem[Kacprzak et al.(2011b)]{kacprzak11b} Kacprzak, G.~G., Churchill, C.~W., Evans, J.~L., Murphy, M.~T., \& Steidel, C.~C.\ 2011b, \mnras, 416, 3118  
\bibitem[Kacprzak et al.(2012a)]{kacprzak12a} Kacprzak, G.~G., Churchill, C.~W., \& Nielsen, N.~M.\ 2012, \apjl, 760, L7  
\bibitem[Kacprzak et al.(2012b)]{kacprzak12b} Kacprzak, G.~G., Churchill, C.~W., Steidel, C.~C., Spitler, L.~R., \& Holtzman, J.~A.\ 2012, \mnras, 427, 3029 
\bibitem[Kacprzak et al.(2008)]{kacprzak08} Kacprzak, G.~G., Churchill, C.~W., Steidel, C.~C., \& Murphy, M.~T.\ 2008, \aj, 135, 922-927 
\bibitem[Kacprzak et al.(2014)]{kacprzak14} Kacprzak, G.~G., Martin, C.~L., Bouch{\'e}, N., et al.\ 2014, \apjl, 792, L12 
\bibitem[Kacprzak et al.(2010b)]{kacprzak10b} Kacprzak, G.~G., Murphy, M.~T., \& Churchill, C.~W.\ 2010, \mnras, 406, 445 
\bibitem[Kacprzak et al.(2015)]{kacprzak15} Kacprzak, G.~G., Muzahid, S., Churchill, C.~W., Nielsen, N.~M., \& Charlton, J.~C.\ 2015, \apj, 815, 22 
\bibitem[Kacprzak et al.(2016)]{kacprzak16} Kacprzak, G.~G., van de Voort, F., Glazebrook, K., et al.\ 2016, \apjl, 826, L11
\bibitem[Keeney et al.(2013)]{keeney13} Keeney, B.~A., Stocke, J.~T., Rosenberg, J.~L., et al.\ 2013, \apj, 765, 27 
\bibitem[Kere{\v s} et al.(2005)]{keres05} Kere{\v s}, D., Katz, N., Weinberg, D.~H., \& Dav{\'e}, R.\ 2005, \mnras, 363, 2 
\bibitem[Kere{\v s} et al.(2009)]{keres09} Kere{\v s}, D., Katz, N., Fardal, M., Dav{\'e}, R., \& Weinberg, D.~H.\ 2009, \mnras, 395, 160 
\bibitem[Kimm et al.(2011)]{kimm11} Kimm, T., Slyz, A., Devriendt, J., \& Pichon, C.\ 2011, \mnras, 413, L51 
\bibitem[Krogager et al.(2013)]{krogager13} Krogager, J.-K., Fynbo, J.~P.~U., Ledoux, C., et al.\ 2013, \mnras, 433, 3091 
\bibitem[Kunth \& Bergeron(1984)]{kunth84} Kunth, D., \& Bergeron, J.\ 1984, \mnras, 210, 873 
\bibitem[Lan et al.(2014)]{lan14} Lan, T.-W., M{\'e}nard, B., \& Zhu, G.\ 2014, \apj, 795, 31 
\bibitem[Lanzetta \& Bowen(1992)]{lanzetta92} Lanzetta, K.~M., \& Bowen, D.~V.\ 1992, \apj, 391, 48 
\bibitem[Lehner et al.(2013)]{lehner13} Lehner, N., Howk, J.~C., Tripp, T.~M., et al.\ 2013, \apj, 770, 138 
\bibitem[Lehner et al.(2016)]{lehner16} Lehner, N., O'Meara, J.~M., Howk, J.~C., Prochaska, J.~X., \& Fumagalli, M.\ 2016, ApJ, submitted, arXiv:1608.02588 
\bibitem[Lehner et al.(2009)]{lehner09} Lehner, N., Prochaska, J.~X., Kobulnicky, H.~A., et al.\ 2009, \apj, 694, 734 
\bibitem[Liang \& Chen(2014)]{liang14} Liang, C.~J., \& Chen, H.-W.\ 2014, \mnras, 445, 2061 
\bibitem[Martin et al.(2012)]{martin12} Martin, C.~L., Shapley, A.~E., Coil, A.~L., et al.\ 2012, \apj, 760, 127 
\bibitem[Martin et al.(2015)]{martin15} Martin, D.~C., Matuszewski, M., Morrissey, P., et al.\ 2015, \nat, 524, 192 
\bibitem[Martin et al.(2016)]{martin16} Martin, D.~C., Matuszewski, M., Morrissey, P., et al.\ 2016, \apjl, 824, L5 
\bibitem[Mathes et al.(2014)]{mathes14} Mathes, N.~L., Churchill, C.~W., Kacprzak, G.~G., et al.\ 2014, \apj, 792, 128 
\bibitem[Meiring et al.(2013)]{meiring13} Meiring, J.~D., Tripp, T.~M., Werk, J.~K., et al.\ 2013, \apj, 767, 49 
\bibitem[Muzahid et al.(2016)]{muzahid16} Muzahid, S., Kacprzak, G.~G., Charlton, J.~C., \& Churchill, C.~W.\ 2016, \apj, 823, 66 
\bibitem[Muzahid et al.(2015)]{muzahid15} Muzahid, S., Kacprzak, G.~G., Churchill, C.~W., et al.\ 2015, \apj, 811, 132
\bibitem[Navarro et al.(1996)]{nfw96} Navarro, J.~F., Frenk, C.~S., \& White, S.~D.~M.\ 1996, \apj, 462, 563 
\bibitem[Nielsen et al.(2013a)]{nielsen13a} Nielsen, N.~M., Churchill, C.~W., Kacprzak, G.~G., \& Murphy, M.~T.\ 2013, \apj, 776, 114 
\bibitem[Nielsen et al.(2013b)]{nielsen13b} Nielsen, N.~M., Churchill, C.~W., \& Kacprzak, G.~G.\ 2013, \apj, 776, 115 
\bibitem[Nielsen et al.(2015)]{nielsen15} Nielsen, N.~M., Churchill, C.~W., Kacprzak, G.~G., Murphy, M.~T., \& Evans, J.~L.\ 2015, \apj, 812, 83 
\bibitem[Nielsen et al.(2016)]{nielsen16} Nielsen, N.~M., Churchill, C.~W., Kacprzak, G.~G., Murphy, M.~T., \& Evans, J.~L.\ 2016, \apj, 818, 171 
\bibitem[Ocvirk et al.(2008)]{ocvirk08} Ocvirk, P., Pichon, C., \& Teyssier, R.\ 2008, \mnras, 390, 1326
\bibitem[Oppenheimer et al.(2012)]{oppenheimer12} Oppenheimer, B.~D., Dav{\'e}, R., Katz, N., Kollmeier, J.~A., \& Weinberg, D.~H.\ 2012, \mnras, 420, 829 
\bibitem[Penprase et al.(2010)]{penprase10} Penprase, B.~E., Prochaska, J.~X., Sargent, W.~L.~W., Toro-Martinez, I., \& Beeler, D.~J.\ 2010, \apj, 721, 1 
\bibitem[P{\'e}roux et al.(2011)]{peroux11} P{\'e}roux, C., Bouch{\'e}, N., Kulkarni, V.~P., York, D.~G., \& Vladilo, G.\ 2011, \mnras, 410, 2237 
\bibitem[P{\'e}roux et al.(2016)]{peroux16} P{\'e}roux, C., Rahmani, H., Quiret, S., et al.\ 2016, arXiv:1609.07389 
\bibitem[Prochaska et al.(2003)]{prochaska03} Prochaska, J.~X., Gawiser, E., Wolfe, A.~M., Cooke, J., \& Gelino, D.\ 2003, \apjs, 147, 227 
\bibitem[Prochaska et al.(2011)]{prochaska11} Prochaska, J.~X., Weiner, B., Chen, H.-W., Mulchaey, J., \& Cooksey, K.\ 2011, \apj, 740, 91 
\bibitem[Prochaska \& Wolfe(1997)]{prochaska97} Prochaska, J.~X., \& Wolfe, A.~M.\ 1997, \apj, 487, 73 
\bibitem[Quiret et al.(2016)]{quiret16} Quiret, S., P{\'e}roux, C., Zafar, T., et al.\ 2016, \mnras, 458, 4074 
\bibitem[Rafelski et al.(2012)]{rafelski12} Rafelski, M., Wolfe, A.~M., Prochaska, J.~X., Neeleman, M., \& Mendez, A.~J.\ 2012, \apj, 755, 89 
\bibitem[Rahmani et al.(2016)]{rahmani16} Rahmani, H., P{\'e}roux, C., Turnshek, D.~A., et al.\ 2016, \mnras, 463, 980 
\bibitem[Ribaudo et al.(2011)]{ribaudo11} Ribaudo, J., Lehner, N., Howk, J.~C., et al.\ 2011, \apj, 743, 207 
\bibitem[Richter(2012)]{richter12} Richter, P.\ 2012, \apj, 750, 165 
\bibitem[Richter et al.(2016)]{richter16} Richter, P., Wakker, B.~P., Fechner, C., et al.\ 2016, \aap, 590, A68 
\bibitem[Rubin et al.(2012)]{rubin12} Rubin, K.~H.~R., Prochaska, J.~X., Koo, D.~C., \& Phillips, A.~C.\ 2012, \apjl, 747, L26 
\bibitem[Rudie et al.(2012)]{rudie12} Rudie, G.~C., Steidel, C.~C., Trainor, R.~F., et al.\ 2012, \apj, 750, 67 
\bibitem[Savage et al.(2003)]{savage03} Savage, B.~D., Sembach, K.~R., Wakker, B.~P., et al.\ 2003, \apjs, 146, 125 
\bibitem[Schroetter et al.(2015)]{schroetter15} Schroetter, I., Bouch{\'e}, N., P{\'e}roux, C., et al.\ 2015, \apj, 804, 83 
\bibitem[Schroetter et al.(2016)]{schroetter16} Schroetter, I., Bouch{\'e}, N., Wendt, M., et al.\ 2016, arXiv:1605.03412
\bibitem[Sembach et al.(2004)]{sembach04} Sembach, K.~R., Tripp, T.~M., Savage, B.~D., \& Richter, P.\ 2004, \apjs, 155, 351
\bibitem[Shen et al.(2012)]{shen12} Shen, S., Madau, P., Aguirre, A., et al.\ 2012, \apj, 760, 50 
\bibitem[Shen et al.(2013)]{shen13} Shen, S., Madau, P., Guedes, J., et al.\ 2013, \apj, 765, 89 
\bibitem[Steidel(1995)]{s95} Steidel, C.~C.\ 1995, QSO Absorption Lines, 139 
\bibitem[Steidel et al.(1994)]{sdp94} Steidel, C. C., Dickinson, M., \& Persson, S. E.  1994, ApJ, 437, L75
\bibitem[Steidel et al.(2010)]{steidel10} Steidel, C.~C., Erb, D.~K., Shapley, A.~E., et al.\ 2010, \apj, 717, 289 
\bibitem[Steidel et al.(2002)]{steidel02} Steidel, C.~C., Kollmeier, J.~A., Shapley, A.~E., et al.\ 2002, \apj, 570, 526 
\bibitem[Stewart et al.(2013)]{stewart16} Stewart, K.~R., Brooks, A.~M., Bullock, J.~S., et al.\ 2013, \apj, 769, 74 
\bibitem[Stewart et al.(2011)]{stewart11} Stewart, K.~R., Kaufmann, T., Bullock, J.~S., et al.\ 2011, \apj, 738, 39 
\bibitem[Stewart et al.(2016)]{stewart13} Stewart, K., Maller, A., O{\~n}orbe, J., et al.\ 2016, arXiv:1606.08542 
\bibitem[Stocke et al.(2010)]{stocke10} Stocke, J.~T., Keeney, B.~A., \& Danforth, C.~W.\ 2010, \pasa, 27, 256 
\bibitem[Stocke et al.(2013)]{stocke13} Stocke, J.~T., Keeney, B.~A., Danforth, C.~W., et al.\ 2013, \apj, 763, 148 
\bibitem[Stocke et al.(2006)]{stocke06} Stocke, J.~T., Penton, S.~V., Danforth, C.~W., et al.\ 2006, \apj, 641, 217
\bibitem[Thom et al.(2011)]{thom11} Thom, C., Werk, J.~K., Tumlinson, J., et al.\ 2011, \apj, 736, 1 
\bibitem[Tripp et al.(2005)]{tripp05} Tripp, T.~M., Jenkins, E.~B., Bowen, D.~V., et al.\ 2005, \apj, 619, 714
\bibitem[Tripp et al.(1998)]{tripp98} Tripp, T.~M., Lu, L., \& Savage, B.~D.\ 1998, \apj, 508, 200 
\bibitem[Tripp et al.(2008)]{tripp08} Tripp, T.~M., Sembach, K.~R., Bowen, D.~V., et al.\ 2008, \apjs, 177, 39
\bibitem[Tumlinson et al.(2011)]{tumlinson11} Tumlinson, J., Thom, C., Werk, J.~K., et al.\ 2011, Science, 334, 948
\bibitem[van de Voort \& Schaye(2012)]{vandevoort12} van de Voort, F., \& Schaye, J.\ 2012, \mnras, 423, 2991 
\bibitem[van de Voort et al.(2011)]{vandevoort11} van de Voort, F., Schaye, J., Booth, C.~M., Haas, M.~R., \& Dalla Vecchia, C.\ 2011, \mnras, 414, 2458 
\bibitem[Wakker \& Savage(2009)]{wakker09} Wakker, B.~P., \& Savage, B.~D.\ 2009, \apjs, 182, 378
\bibitem[Wolfe et al.(2005)]{wolfe05} Wolfe, A.~M., Gawiser, E., \& Prochaska, J.~X.\ 2005, \araa, 43, 861 
\bibitem[Wotta et al.(2016)]{wotta16} Wotta, C.~B., Lehner, N., Howk, J.~C., O'Meara, J.~M., \& Prochaska, J.~X.\ 2016, arXiv:1608.02584 
\bibitem[Yoshida et al.(2004)]{yoshida04} Yoshida, N., Bromm, V., \& Hernquist, L.\ 2004, \apj, 605, 579 
\bibitem[Zheng et al.(2015)]{zheng15} Zheng, Y., Putman, M.~E., Peek, J.~E.~G., \& Joung, M.~R.\ 2015, \apj, 807, 103 
\bibitem[Zibetti et al.(2007)]{zibetti07} Zibetti, S., M{\'e}nard, B., Nestor, D.~B., et al. 2007, ApJ, 658, 161
 
\end{thebibliography}
